
\magnification=1200
\def\ni{\noindent}
\def\.{\mathaccent 95}
\def\a{\alpha}
\def\be{\beta}

\def\Om{\Omega}

\def\frac#1#2{{\textstyle{{#1}\over {#2}}}}
\def\ni{\noindent}
\def\lsim{\mathrel{\rlap{\lower4pt\hbox{\hskip1pt$\sim$}}
    \raise1pt\hbox{$<$}}}
\def\gsim{\mathrel{\rlap{\lower4pt\hbox{\hskip1pt$\sim$}}
    \raise1pt\hbox{$>$}}}
\def\sqr#1#2{{\vcenter{\vbox{\hrule height.#2pt
         \hbox{\vrule width.#2pt height#1pt \kern#1pt
         \vrule width.#2pt}
         \hrule height.#2pt}}}}

\newbox\grsign \setbox\grsign=\hbox{$>$} \newdimen\grdimen \grdimen=\ht\grsign
\newbox\simlessbox \newbox\simgreatbox
\setbox\simgreatbox=\hbox{\raise.5ex\hbox{$>$}\llap
     {\lower.5ex\hbox{$\sim$}}}\ht1=\grdimen\dp1=0pt
\setbox\simlessbox=\hbox{\raise.5ex\hbox{$<$}\llap
     {\lower.5ex\hbox{$\sim$}}}\ht2=\grdimen\dp2=0pt

%
%

\def\ref#1  {\noindent \hangindent=24.0pt \hangafter=1 {#1} \par}
\def\doublespace {\smallskipamount=6pt plus2pt minus2pt
                  \medskipamount=12pt plus4pt minus4pt
                  \bigskipamount=24pt plus8pt minus8pt
                  \normalbaselineskip=24pt plus0pt minus0pt
                  \normallineskip=2pt
                  \normallineskiplimit=0pt
                  \jot=6pt
                  {\def\smallskip {\vskip\smallskipamount}}
                  {\def\medskip   {\vskip\medskipamount}}
                  {\def\bigskip   {\vskip\bigskipamount}}
                  {\setbox\strutbox=\hbox{\vrule 
                    height17.0pt depth7.0pt width 0pt}}
                  \parskip 12.0pt
                  \normalbaselines}
\def\ts{\times}
\def\lb{\langle}
\def\rb{\rangle}
\def\curl{\nabla {\ts}}

\def\bbv{\bar {\bf v}}
\def\bfvp{{\bf v}'}
\def\bfjp{{\bf j}'}
\def\bfv{{\bf v}}

\def\bfwp{{\bomega}'}
\def\bfBp{{\bf B}'}

\def\nb{\nabla}
\def\curl{\nb\ts}

\def\b0{b'^{(0)}}
\def\v0{v'^{(0)}}
\def\w0{\omega'^{(0)}}
\def\bb0{\bfbp^{(0)}}
\def\bv0{\bfvp^{(0)}}
\def\bw0{\bfwp^{(0)}}
\def\bj0{\bfjp^{(0)}}

\def\bbB{\bar {\bf B}}
\def\bfB{\bf B}

\doublespace

\centerline{IN SITU ORIGIN OF LARGE SCALE GALACTIC}
\centerline {MAGNETIC FIELDS WITHOUT KINETIC HELICITY?}
\centerline{Eric G. Blackman}
\centerline{Institute of Astronomy, Madingley Road, Cambridge CB3 OHA, 
 England}              
\medskip

\centerline {\bf ABSTRACT}

The origin and sustenance of 
large scale galactic magnetic fields has been a long standing 
and controversial astrophysical problem. Here an alternative to 
the ``standard''  $\a-\Omega$ mean field dynamo and primordial 
theories is pursued.  The steady supply of supernovae 
induced turbulence exponentiates the total field energy, 
providing a significant seed mean field that can be 
linearly stretched by shear.
The observed micro-Gauss fields would be produced primarily within 
one vertical diffusion time since it is only during this time that
linear stretching can compete with diffusion.
This approach does not invoke exponential mean field dynamo
growth from the helicity $\a$-effect but does  
employ turbulent diffusion, which limits the number of large scale reversals.
The approach could be of interest if the non-linear Galactic 
dynamo helicity effect is suppressed independently of the turbulent
diffusion.  This is an important, but presently unresolved issue.

\medskip

\centerline{{\bf Subject Headings}: Galaxy: general; galaxies: magnetic fields;
 ISM: magnetic fields}

\vfill
\eject

\centerline{\bf 1. Introduction}
\ni Magnetic fields are important to dynamics or emission in
almost all astrophysical systems.  The formation of observed micro-Gauss 
large scale magnetic fields in the interstellar medium (ISM) of spiral 
galaxies has been considered a fundamental unsolved astrophysical problem 
(Beck et al. 1996; Zweibel \& Heiles 1997).  
There is an important 
distinction between the large and small scale magnetic fields observed 
in the ISM.  Observations indicate that the small scale field is typically
ordered on sub-kiloparsec scales and superimposed on that
is a large scale, primarily azimuthal field.  While it is more or less
agreed that the small scale component of this field is injected
and sustained by supernovae turbulence (e.g. Beck et al. 1996), 
the generation of the large scale field is where the
controversy lies.  

Whether or not the large scale fields are of primordial origin 
or are produced in situ is difficult to determine
observationally, and the ``standard'' competing  explanations both suffer
from difficulties.  For example even if primordial fields could be 
produced, in situ processing of the field during the galactic lifetime
needs to be addressed, and would likely dominate the present character of the 
observed field.  The standard in situ 
mean field ``$\a-\Om$ dynamo'' (Parker 1979; Moffatt 1978; Ruzmaikin et al. 1988) has been threatened by some non-linear simulations (e.g. Tao et al. 1993; 
Cattaneo \& Hughes 1996).
While stretching of field lines  
by differential rotation (which provides the ``$\Omega$-effect'') is 
not controversial, the helical property of the turbulence 
(which provides the ``$\a$-effect'') and the turbulent 
diffusion (the ''$\beta$ effect'') may be suppressed.
Here I suppose that the 
$\a$-effect is suppressed but that the $\beta$ effect is not.
Whether these two effects can be disentangled is unclear
(Field \& Blackman 1997) but the possibility can still be explored.

Turbulent amplification of 
the small scale field can provide a steadily supplied 
seed mean field (MF) whose subsequent stretching by 
differential rotation may produce and sustain large scale 
Galactic magnetic fields without  a 
dynamo $\a$-effect or a primordial seed.
I first give the basic magnetic MF equations and 
summarize the complications of standard MF dynamo theory. 
The MF equation is then solved without the
 $\a$-effect, but keeping in turbulent diffusion.
It is shown that in 1 vertical diffusion time, differential shearing
can increase the MF by an order of magnitude.  
It is also shown that the MF is maximized on a radial scale
of a kpc or so: Turbulent diffusion tends to favor large scale
field stretching, but this competes with the  
seed field's inverse dependence on scale.
Simple statistical arguments are then used to predict the most likely
number of reversals.  
That the small scale field is not necessarily dominant on only one
scale is also emphasized.
The assumptions employed here are no more controversial
than in standard MF theory.  Generally speaking, MF theory as applied to
the Galaxy is most certainly an oversimplification, but
it does provide a useful framework from which some
understanding can be gained.

\centerline {\bf 2. Aspects of Standard Theory}
Writing the magnetic field (and velocity) as a sum of mean
and fluctuating components, ie. ${\bf B} = {\bar {\bf B}}+{\bf B}'$, 
the MF equation can be obtained by spatially averaging the magnetic
induction equation over scales large
compared to the fluctuations but small compared to the
overall scale of the system (e.g. galaxy):  
$$\partial_{t}{\bbB} =\curl(\bbv\ts\bbB) +\curl\lb\bfvp\ts\bfBp\rb 
+\nu_M\nabla^{2}{\bbB}, \eqno(1)$$
where $\bfv$ is the velocity and $\nu_M$ is the magnetic diffusivity,
and the primed and barred (or  bracketed) 
quantities indicate fluctuating and mean
values respectively.  
The turbulent EMF in (1) can be written
$\lb\bfvp\ts\bfBp\rb=\a_{ij}({\bfB},{\bfv}){\bar B}_j-\beta_{ijk}({\bfB},{\bfv})
\nabla_j{\bar B}_k+\ ...$
where ``...'' indicates higher order gradients.
In linear kinematic dynamo theory for
isotropic incompressible plasmas, 
$\a\propto\lb \bfvp\cdot \int\curl \bfvp(t')dt'\rb$ and
$\beta\propto\lb \bfvp\cdot \int \bfvp(t')dt'\rb$.
In the dynamic, non-linear theory, these can be functions of
$\bf B$,  and are not necessarily isotropic.

In the standard $\a-\Om$ dynamo $(\a\Om D)$  applied to the Galaxy  
(e.g. Ruzmaikin et al. 1988),  supernovae (SN) induced turbulent eddies on
 $\sim 100$pc scales stretch field lines into loops or cells. 
The coriolis force, in principle, conspires to 
statistically twist all of these loops in the same 
direction, providing a much larger scale mean
loop.  This, resulting from the large scale 
reflection asymmetry, provides the non-vanishing pseudoscalar
$\a$-effect (Parker 1979; Moffatt 1978).  
The outer portions of these loops must incur turbulent diffusion 
to leave a net  mean flux in each hemisphere of the Galactic disk 
(Parker 1979).  The large scale field formed in this way is further
sheared by differential rotation (the $\Omega$-effect), 
providing new toroidal field and
starting the process again.   In principle, this feedback
leads to exponential MF growth
of a primarily azimuthal field with a growth time $\sim 4\ts 10^8$ yr (Ruzmaikin et al. 1988).

Standard kinematic $\a\Om D$ treatments ignore the
backreaction of the growing magnetic field on the turbulence. 
Because high magnetic Reynolds numbers
make the last term in (1) negligible on the energy containing scales, 
the field exponentially grows to 
equipartition with the turbulent energy by a fast dynamo (FD) 
on a time scale much shorter than any MF evolution time (e.g. Parker 1979)
and does not require helicity.
In combination with even a weak MF that
is $\gsim \rho^{1/2} v'/R_m^{1/2}$ (where $R_m$ is the magnetic
Reynolds number, and $\rho$ is the density), this may make 
Lorentz forces (Kulsrud \& Anderson 1992; Cattaneo 1994)
lock a significant fraction of motions into oscillations.
The magnetic fields act like springs that the 
required turbulent motions must fight against.  
Although simulations in 2-D show $\beta$ suppression (Cattaneo 1994), 
there are not yet simulations that show $\beta$ suppression in 3-D.
There have been some simulations showing 
$\alpha$-effect suppression in 3-D  
(Tao et al., 1993; Cattaneo \& Hughes 1996), and others that do not 
(Brandenburg \& Donner 1997).
Intermittency (Blackman 1996; Subramanian 1997) and the nature of the
forcing function may 
play a role in overcoming both $\alpha$ and $\beta$ suppression in 
3-D. Basically, there is no clear consensus on what happens in fully
non-linear mean field dynamo theory with respect to the backreaction, even 
as to whether suppression of $\a$ and $\beta$ are intertwined 
(e.g. Field \& Blackman 1997) or independent.


\centerline {\bf 3. Linear Mean Field Growth from a Random Seed Field}
\ni SN inject turbulent energy into the ISM, and
also inject magnetic field (Ruzmaikin et al. 1988; Rees 1994).  
Because of the observed dispersal of heavy elements (Rana 1991), the supernova
ejecta at least mix with the remnant material.  
Theoretical estimates for the mean seed field injected from SN 
range from $10^{-13}$ Gauss from simple flux freezing, to $10^{-8}$ Gauss 
for including winding in pulsar winds (Rees 1994). 

Both theory and simulation (Parker 1979; Piddington 1992; Beck et al. 1996)
show that the FD builds up small scale 
magnetic field energy on a growth time of order the energy containing eddy
turnover time $\sim l/v' \sim 10^7$ yr, where 
$l\sim 100$pc is usually taken as the energy containing 
eddy scale and $v'\gsim 10$km/sec is a typical observed speed of
these eddies.  Thus, once the Galactic volume is full of
a weak seed magnetic field from the first set of SN, 
the next generation of eddies stirs the field to equipartition.
The SN remnants fill the Galactic disk (height $\sim 500$pc by radius $12$kpc)  
every $10^7$ years given their observed rate
of $\sim 0.02{\rm yr^{-1}}$ (Ruzmaikin et al. 1988), and this maintains 
a steady random field energy. 
How the field actually mixes from the SN
to the ambient ISM is  complicated. 
The amount of magnetic annihilation, the amount of enhancement,
the geometry/topology of the injected field (e.g. Ruzmaikin et al. 1988), 
and the role of boundary instabilities are all subtle issues.
Despite these complications, the basic
picture of SN seed field injection and subsequent stirring as
described above, leading to 
equipartition fields at a time $\lsim 2.5 \ts 10^8$yr
(where this upper limit comes from using the $10^{-13}$G seed value
given above) and all subsequent times, is consistent 
with observations:
The magnitude of the random field is observed 
to be $B\sim 5-10\ts 10^{-6}$ G (e.g. Heiles 1994; Rand \& Kulkarni 1989).
Rand \& Kulkarni (1989) impose a single cell model
whose best fit small scale over which the field is ordered
is 50-100pc.
Though this has become the standard quoted range for the small scale,
it will be emphasized later why  multiple and larger
cell sizes (Ferri\`ere 1994) are important.

The above small scale field would give a corresponding  
MF of  magnitude 
${\bar B}_{0}\sim B/N^{1/2}$
where 
$N$ is the number of small scale
coherence volumes in the region of averaging.
Such a residual large scale  field 
has been argued to be a viable source of seed field for the 
Galactic $\a\Om$D (Ruzmaikin et al. 1988; Rees 1994),
but below I suggest that even if $\a$ is suppressed, 
an appropriate large scale field can still be produced.  
Assume $\a$ is suppressed 
well below the critical value required (Parker 1979)
for the standard $\a-\Omega$ dynamo
growth, ie., $\a << \a_{crit}\sim \beta^2/(\Om h^3)$ (where
$h$ is the disk height), so we can then ignore it in what follows. 
The MF induction equations in cylindrical coordinates become

$$\partial_t{\bar B_{r}} =
\be\nabla^2{\bar B}_{r}\eqno(2)$$

$$\partial_t{\bar B_{\phi}} 
=r\partial_r\Om{\bar B}_{r}+\beta \nabla^2 {\bar B_{\phi}},\eqno(3)$$

$$\partial_t{\bar B_z} =
\beta \nabla^2 {\bar B}_z,\eqno(4)$$
Eqs. (2) and (4) are decoupled from (3),  implying pure 
diffusion of ${\bar B}_r$ and ${\bar B}_z$.
As in standard treatments (e.g. Ruzmaikin et al. 1988), the 
$\Omega$-effect (differential rotation)
increases the azimuthal field in (3) linearly on a time
of order the rotation time. 
For the Galaxy this is $\Omega^{-1}\sim 3.3\ts 10^{7}$yr.   
Because the rotational energy far exceeds that which can be 
transferred into the magnetic field during the age of the universe, 
the $\Om$-effect is not controversial;  there is no back-reaction on 
the large scale rotational motion.  
As shown below, this $\Om$-effect can  generate a factor of 
$\sim 10$ increase in the large scale field.  This is sufficient 
without an $\a$-effect since the seed field is continually
supplied.  
                   
To solve the equations and determine the dominant
scale of the MF, I
assume that  ${\bar {\bfB}}= {\bar {\bf  B}}_{t}e^{(i{\bf k}\cdot {\bf x})}$, where 
$k$ is the wave vector of the MF and the 
subscript $t$ labels the time dependence.  
The MF equations (e.g. Ruzmaikin et al. 1988) without an $\a$-effect and with homogeneous $\beta$
for the azimuthal and radial fields are then

$$\partial_t{\bar B_{\phi t}}
={\bar B}_{r t}f\Omega-\beta k^2 {\bar B_{\phi t}},\eqno(5)$$
and
$$\partial_t {\bar B_{r t}}= -\beta k^2 {\bar B_{r t}},\eqno(6)$$
where $f\Om=r\partial\Omega/\partial r$. 
Solving (6) gives ${\bar B}_{r t}={\bar B}_{r 0}{\rm Exp}[-k^2\beta t]$, so 
(5) gives
$$\partial_t{\bar B}_{\phi t}=f\Om B_{r 0} 
{\rm Exp}[-k^2\beta t] -k^2\beta{\bar B}_{r t}. \eqno(7)$$
Multiplying both sides of (7) by ${\rm Exp}[k^2\beta t]$, using the chain rule
and solving gives
$${\bar B}_{\phi t}= B(f\Om t+1)(3N)^{-1/2}{\rm Exp}[-(k_\phi^2+k_z^2+k_r^2)\beta t]$$
$$\simeq B(f\Om t+1)[(3 hDR)(4\pi l^3/3)]^{-1/2}{\rm Exp}[-(D^{-2}+h^{-2}+R^{-2})\beta t], \eqno(8)$$
where I have taken ${\bar B}_{r 0}\sim {\bar B}_{\phi 0}
\sim {\bar B}_{0}/3^{1/2}=
B/(3N)^{1/2}$, $N\sim h D R/(4\pi l^3/3)$, and 
$h$ is the Galactic scale height, while $D$ 
and $R$ are the azimuthal and radial mean field gradient
lengths corresponding to their wave vectors, and
defined only for scales $\gsim h$.
  
The dominant contribution to the MF at any one time is that
produced within a vertical diffusion time, $\tau_{dv}$,
from the observation time.  Thus 
$\beta t\sim 1/h^2$, and from (8) 
$${\bar B}_{\phi \tau_{dv}} =B(f \Om h^2/\beta+1)
[3 hDR/(4\pi l^3/3)]^{-1/2}{\rm Exp}[-(1+h^2/D^{2}+h^2/R^{2})].\eqno(9)$$
The scale height of the disk is fixed at $h\sim 500$pc, but 
${\bar B}_{\phi \tau_{dv}}$ can be extremized as a function
of $R$ and $D$, giving a maximum at $R=D=2(\be t)^{-1/2}$.  
Using $\beta\sim 10^{26}$ cm$^2$/sec (e.g. Ruzmaikin et al. 1988)
a Galactic disk scale-height of $h \sim 500$pc,  
$\Om\sim 10^{-15}{\rm sec^{-1}}$, and $f\sim 1$ this gives
$D=R=1$kpc.  Thus 
${\bar B}_{\phi \tau_{dv}}= 1.42\ts 10^{-6} (B/5\ts 10^{-6}{\rm G})$G .
If instead we take $D=3$kpc, to match the radial 
scale measured by Faraday rotation (e.g. Rand \& Lyne 1994),
this becomes $1.02\ts 10^{-6}(B/5\ts 10^{-6}{\rm G})$G.
The approximate magnitude of the local MF (e.g. Heiles 1994)
may therefore be reproduced without the $\a$-effect.

Note that ${\bar B}_{\phi \tau_{dv}}$ 
depends on the value of the characteristic
averaging azimuthal distance as $D^{-1/2}{\rm Exp}[-h^2/D^2]$, 
on the characteristic small scale structure size $l$ 
to the $-3/2$ power, and linearly on $B$.
The actual small scale field of the Galaxy 
has been shown from observations  to be inconsistent 
with a single scale size (Rand \& Kulkarni 1989):  
The statistical dispersion between the
observed field and their model large scale field shows little evidence of
fall off with the distance to  pulsars as it should if the 
the single cell size model were appropriate.
This highlights the importance of super-bubbles and other larger
scale fluctuations known to be important to the field structure
(Ferri\`ere, 1996). If the region over which the measured MF  
were composed of primarily $\gsim 200$pc instead of 100pc structures, 
then the estimate given above is magnified
by an additional factor $\gsim 2^{3/2}=2.8$.
Also, Heiles (1994) finds that the total magnetic energy
does not scale simply with the mean azimuthal field as measured by
Faraday rotation in different
parts of the Galaxy.  This can be explained
in the present model, since the mean field is proportional to
the RMS field divided by $N^{1/2}$, where $N$ is the number of
small scale cells of uniform field in the region determining the Faraday
rotation measure. Regions of different cell sizes would therefore produce
different observed mean fields even if the total magnetic energy density 
were the same. All of this highlights the possible importance of
multiple small scale sizes.

\centerline{\bf 4. Discussion of Mean Field Reversals}
\ni The calculation of section 3 shows that a scale height of 
$h=500$pc, maximizes the azimuthal MF for a radial averaging scale of
$D=R=1$kpc. This results from two competing effects: [1]
\ni The time for shear to increase the field strength by an order
of magnitude is relatively independent
of scale.  Thus large scales are preferentially sheared 
in a fixed time because the competing
turbulent diffusion depends on scale squared.  
[2] However, the initial seed field depends inversely on the 
averaging scale.  The scale of 1kpc optimizes [1] and [2].
This defines the minimum radial scale over which the
maximum average azimuthal field could reverse sign. 
This does not mean that there would be necessarily be reversals every 1kpc.
It means that between 1kpc annuli, the 
mean field may or may not reverse. Within a 10kpc Galactic radius, 
there are $\sim $ 9 interfaces between 1kpc annuli. 
The probability $P(n)$ of observing $n$ reversals by Faraday rotation would
be 9 'choose' $n$, i.e. 9!/(9-$n$)!$n$!, which is maximized for $n$=4 or 5. 


Galactic Faraday rotation observations can determine
the sign of the large scale field in the line of sight 
(cf. Beck et al. 1996; Zwiebel \& Heiles 1997). 
(Unlike  Galactic measurements, where pulsar dispersion measures
can be used, extragalactic measurements require independent  determinations
of the density to obtain any information from 
Faraday rotation.  The data for external galaxies are 
therefore less reliable (Heiles 1994; Zwiebel \& Heiles 1997).)
Generally, a large scale theoretical Galactic field model  
is statistically compared to observations (e.g. Rand \& Kulkarni 1989).
Field reversals seem to occur in each of the two interarm regions immediately
inside of the solar circle (Beck et al. 1996; Heiles 1994)
with perhaps two more outside.
The reversals are not necessarily periodic between spiral arms (Vall\'ee 1996).
Also, because of fluctuations in  rotation measure data 
for some quadrants (Rand \& Lyne 1994; Beck et al. 1996)
averaging over smaller scales then
shows smaller intermediate scale reversals.  This
again highlights that intermediate scales (Rand \& Kulkarni 1989) 
from  50-500 pc complicate
theoretical and observational interpretations.  
The precise structure of the large scale field in spiral
galaxies is difficult to conclusively determine (Beck et al. 1996).



Note that turbulent diffusion is distinct from dissipation.  
The former describes a transfer of magnetic energy 
between scales, whereas dissipation is a removal of magnetic energy.
Turbulent motions on sub-kiloparsec scales both
randomize the mean field and amplify 
the small scale field, thereby re-seeding the mean field.  
Though a turbulent cascade drains energy 
to the dissipation scale, the  magnetic energy 
is  steadily replenished by the FD
and the total magnetic energy density remains steady.

Previous work has recognized the importance of
diffusion for reversal reduction (Poezd et al. 1993).  In fact, the 
weaker the $\a$-effect in dynamo models, the 
less vigorously the $\a\Om D $ can compete with turbulent diffusion and
the fewer reversals that survive.  
Primordial models 
are sometimes employed with the assumption that turbulent diffusion
is not operating (Zweibel \& Heiles 1997). 
Though this seems unlikely, 
other proposed mechanisms would then be needed to eliminate reversals 
(c.f. Zweibel \& Heiles 1997).  Another possibility is that 
the winding of a proto-galactic field in the subsequently 
formed galaxy (Howard \& Kulsrud 1997) generates the correct number of 
reversals.




For some external galaxies (e.g. NGC6946), observations indicate that 
the large scale field is actually stronger in the interarm regions 
(Beck \& Hoernes 1996). 
In the present approach, the deficit of large scale field in the spiral
arms would be the result of a reduced shear there (Elmgreen, 1994) and thus an
$f<1$ in Eq. (5).  This is generally consistent with 
rotation curves of NGC6946 and other galaxies which show reduced differential 
rotation in spiral arms (Sofue 1986; Rubin et al. 1980).
In contrast, an enhanced MF strength might result 
in the arms if their increased
electron density dominates the effect of reduced shear.  
The total magnetic energy can be larger in spiral arms
if the turbulent energy is higher there.
Varying density  complicates the 
interpretation of rotation measures of external galaxies
if the density variation cannot be independently measured.

\centerline{\bf 5. Discussion}
\ni It is important to understand whether $\a$ and $\beta$ can 
actually be disentangled.  If so, the main point herein is to suggest that 
it may not be absolutely certain that the observed large scale 
Galactic magnetic field requires a dynamo $\a$-effect 
even if the mean field is produced in situ.
The well-known exponential growth of small scale 
field by the FD and its steady replenishing of seed
MF, combined with the
 subsequent linear growth of large scale azimuthal field by the 
$\Omega$-effect  might supply a large scale Galactic field without 
requiring the dynamo $\alpha$-effect.  
The linear growth may be  sufficient because it
proceeds faster than the time for the field to diffuse below micro-Gauss 
values.  The observed field of any spiral galaxy would be
that produced within $\tau_{dv}$ of the time of observation, and 
since the field is steadily replenished, 
this statement is true at any time in a galaxy's 
lifetime $\ge 10^8$yr from the time of the galaxy's birth.
The most likely  number of reversals in the large 
scale field  within 10kpc radius 
would be of order 4-5 in the presence of turbulent diffusion
for the simplest approach.  (Unlike the cellular model of 
Michel \& Yahil (1973), here turbulence is important.)
If flux tubes were present with a small volume filling fraction, and/or if the 
the energy containing small scale of the field were much reduced from
the semi-empirically determined 100pc scale, 
too many mean-field reversals might be produced by the present approach.
A small filling fraction may also aid the 
standard dynamo (Blackman 1996; Subramanian 1997), 
making the approach herein less useful. 
However, unlike the Sun, the tube filling fraction in 
galaxies may be large if the average particle pressure does not 
overwhelm the magnetic pressure (Blackman 1996).

\ni {\bf Acknowledgements}: Thanks to G. Field and E. Zweibel for discussions,
and to the Aspen Center for Physics.

\noindent  Beck, R.,  et al., 1996, ARA\&A, {\bf 34}, 155.


\ni  Beck, R.,\& Hoernes, P., Nature, 1996, {\bf 379}, 47.

\ni  Blackman, E.G., PRL, 1996, {\bf 77}, 2694.

\ni Brandenburg, A. \& Donner, K.J., 1997, 
MNRAS {\bf 288}, L29.

\noindent Cattaneo, F., ApJ, 1994, $\bf 434$, 200.

\noindent  Cattaneo, F. \& Hughes, D.W., 1996, submitted to Phys.Rev.E.

\ni  Elmgreen, B. ApJ, 1994, {\bf 433} 39.

\ni Ferri\`ere, K., 1996, A\&A {\bf 310}, 438.

\ni Field, G.B. \& Blackman, E.G., 1997, preprint.

\noindent  Heiles C., 1994,  in {\it Physics of the Interstellar and 
Intergalactic Medium},   ed.by  A. Ferrara et al. (San Francisco: Astronomical Society of the Pacific).

\ni Howard, A.M. \& Kulsrud, R.M. 1997, ApJ, $\bf 483$, 648. 

\ni Kulsrud R.M., \&  Anderson, S.W., 1992,  ApJ, $\bf 396$, 606.



\ni  Michel F.C., 1973, \& Yahil A. ApJ {\bf 179} 771.

\noindent   Moffat, H.K., 1978, {\it Magnetic Field Generation in
Electrically Conducting Fluids}, (Cambridge:  Cambridge Univ. Press).

\ni  Ohno, H., \& Shibata, S., 1993, MNRAS, {\bf 262}, 953.

\noindent  Parker, E.N., 1979, {\it Cosmical Magnetic Fields}, (Oxford:
Clarendon Press).

\noindent  Piddington, J.H., 1981, {\it Cosmical Magnetic Fields}, 
(Malbar:  Krieger).  

\ni  Poezd, A., Shukarvov, A. \& Sokoloff, D., MNRAS, {\bf 264}, 285 (1993).

\noindent  Pouquet, A., Frisch, U, \& L\'eorat, J., 1976, J. Fluid Mech.
{\bf 77}, 321.

\noindent  Rana, N.C., 1991, ARA\&A, {\bf 29}, 129.

\ni  Rand, R.J. \& Kulkarni, S. R., 1989, ApJ, {\bf 343}, 760.

\ni  Rand, R.J. \& Lyne, A.G., MNRAS, 1994, {\bf 268}, 497.

\ni  Raymond, J.C., ApJ, 1992, {\bf 384}, 502.

\noindent  Rees, M.J., 1994,  in {\it Cosmical Magnetism}, edited by
D. Lynden-Bell, (Dodrecht:  Kluwer).

\ni  Rubin, V.C. et al., 1980, ApJ, {\bf 238}, 471. 

\noindent   Ruzmaikin, A.A.,  Shukurov, A.M., 1988,  \&  Sokoloff, D.D.{\it
Magnetic Fields of Galaxies}, (Dodrecht: Kluwer).

\ni Sofue, Y., ApJ, 1996, {\bf 458}, 120.

\ni Subramanian, K., 1997, preprint, submitted to MNRAS, astro-ph/9707280.


\ni Tao, L.,  Cattaneo, F.,  \&  Vainshtein, S.I., 1993, in {\it Solar and
Planetary Dynamos} edited by M.R.E. Proctor et al., (Cambridge:
Cambridge University Press).

\ni  Vall\'ee, J.P., 1996, A\&A, {\bf 308}, 433.


\noindent  Zweibel, E.G. \& Heiles, C., 1997, Nature, {\bf 385} 131.

\end